\documentclass[letterpaper]{article}
\usepackage{modified-aaai}
\usepackage{times}
\usepackage{helvet}
\usepackage{courier}
\usepackage{tabu}
\usepackage{tabularx}
\usepackage{longtable}
\usepackage{graphicx}
\usepackage{hyperref}
\usepackage{float}
\title{Hindsight Analysis of the Chicago Food Inspection Forecasting Model}
\author{Vinesh Kannan, Matthew A. Shapiro, \and Mustafa Bilgic\\
Illinois Institute of Technology\\
v@hawk.iit.edu, shapiro@iit.edu, mbilgic@iit.edu \\
Chicago, Illinois 60616\\
}

\usepackage{xcolor}

\usepackage{amsthm}
\theoremstyle{plain}
\newtheorem{finding}{Finding}

\usepackage{caption}
\usepackage{subcaption}

\usepackage{enumerate}
\usepackage{enumitem}

\newenvironment{packed_enum}{
\begin{enumerate}[topsep=.07cm,leftmargin=0.13in]
  \setlength{\itemsep}{1pt}
  \setlength{\parskip}{0pt}
  \setlength{\parsep}{0pt}
}{\end{enumerate}}

\newenvironment{packed_item}{
\begin{itemize}[topsep=.07cm,leftmargin=0.13in]
   \setlength{\itemsep}{1pt}
   \setlength{\parskip}{0pt}
   \setlength{\parsep}{0pt}
}{\end{itemize}}

\hypersetup{draft}
\begin{document}
\maketitle

\begin{abstract}
\begin{quote}
The Chicago Department of Public Health (CDPH) conducts routine food inspections of over 15,000 food establishments to ensure the health and safety of their patrons. In 2015, CDPH deployed a machine learning model to schedule inspections of establishments based on their likelihood to commit critical food code violations. The City of Chicago released the training data and source code for the model, allowing anyone to examine the model. We provide the first independent analysis of the model, the data, the predictor variables, the performance metrics, and the underlying assumptions. We present a summary of our findings, share lessons learned, and make recommendations to address some of the issues our analysis unearthed.
\end{quote}
\end{abstract}

\section{Introduction}
\label{sec.intro}

The Chicago Department of Public Health (CDPH) conducts routine, unannounced food inspections, also called canvass inspections, of over 15,000 retail food establishments to ensure their compliance with the Chicago food code. Inspections ensure the immediate safety of restaurant customers and employees \cite{jones04a}. Inspections are also considered an important proactive measure in preventing food-borne illness outbreaks \cite{irwin89a}. Furthermore, inspections investigate restaurant safety beyond perceptions of cleanliness and help the public learn which restaurants adhere to public health regulations \cite{jones04a}.

CDPH inspects most food establishments twice a year, with other establishments deemed as lower risk inspected one a year or once every other year \cite{schenkjr15a}. In addition to routine canvass inspections, CDPH also conducts license inspections for newly-opened businesses, complaint inspections in response to submitted concerns or suspected food poisoning, and re-inspections for issues arising from canvass inspections that require correction.

In 2015, CDPH released the results of a partnership with the Chicago Department of Innovation and Technology and data scientists from Civic Consulting Alliance. This joint effort led to a machine learning model to predict which restaurants were likely to have critical food code violations in order to catch threats to public health earlier in each inspection cycle. The model produces risk scores for each retail food establishment and the Director of Food Protection considers the risk scores from the model in addition to other expert knowledge and operational constraints and then creates inspection schedules for the CDPH inspectors (``sanitarians''). The City of Chicago released the training data and source code for the model, allowing anyone to examine the model, propose improvements, or adopt it in another city.

CDPH used only canvass inspections in the training and testing sets in order to avoid sources of bias from complaint inspections. In other geographic areas, researchers found that residents are more likely to report food-borne illness in restaurants serving Asian cuisine than those serving American cuisine, potentially skewing risk measurements \cite{irwin89a}. Nevertheless, the purpose of the CDPH model is to prioritize the schedule of canvass inspections rather than changing the frequency of inspections of any specific restaurant or type of restaurant.

In a simulation based on the testing set, the CDPH model found restaurants with critical violations seven days earlier, on average, and found 69\% of restaurants with critical violations in the first half of inspections \cite{schenkjr15a}. 

In this paper, we provide the first independent analysis of the CDPH model and data in detail, providing further transparency into the model and its decision-making process. We investigate the feature set, the model parameters, the performance metrics, and the implicit and explicit assumptions made by the model and CDPH. In addition to presenting our findings in detail, we discuss best practices, challenges, and our recommendations.

The rest of the paper is organized as follows. We provide an overview of the data and the model in Section \ref{sec.model}. We provide our detailed analysis in Section \ref{sec.analysis}. We discuss lessons learned about how artificial intelligence is used for government in Section \ref{sec.discussion} and then conclude.
\section{Data and Model Overview}
\label{sec.model}

In this section, we provide an overview of the Chicago food code, the dataset and the features, and the predictive model. For more details, refer to \cite{schenkjr15a}, describing the methodology used by CDPH.

\subsection{Dataset}

The dataset used for training and evaluating the model consists of records of routine canvas inspections performed by sanitarians and digitized by CDPH. The input to the model is a set of features regarding the retail food establishment, the inspection, and environmental factors. The model output aims to predict whether or not the retail food establishment will be cited for a critical violation during that inspection.

CDPH chose to split the data into training and test sets as follows: 17,075 inspection records from September 2011 through April 2014 are used for training and 1,637 inspection records from September 2014 through October 2014 are used for testing.

Additionally, CDPH publishes records of all food establishment inspections on the City of Chicago data portal \cite{food_inspections_dataset}. The public dataset does not include some of the features present in the training and testing data, but does include data not used in the model, such as the name of the business and the narrative report by the sanitarian that details any violations issued. For our analysis of the model, we retrieved records of 50,462 routine inspections and 17,088 complaint inspections from the public dataset ranging from January 2010 to June 2018. CDPH adopted a new food code in July 2018, so we excluded records beyond that date.

\begin{table}[!hb]
    \centering
    \begin{tabularx}{\columnwidth}{|l|l|l|l|X|}
        \hline
        \textbf{Rank} & \textbf{Code} & \textbf{Count} & \textbf{Rate} & \textbf{Summary} \\
        \hline
        1 & V3 & 4418 & 0.093 & Food Temperature Requirement \\
        2 & V2 & 2233 & 0.047 & Food Storage Facilities \\
        3 & V8 & 1239 & 0.026 & Rinse Cycle Sanitizing Solution \\
        4 & V12 & 1012 & 0.021 & Adequate Hand Washing Facilities \\
        5 & V11 & 895 & 0.019 & Adequate Toilet Facilities \\
        6 & V6 & 826 & 0.017 & Employee Handwashing and Hygiene \\
        7 & V9 & 307 & 0.006 & Connection to City Water Supply \\
        8 & V10 & 233 & 0.005 & Sewage and Waste Water Disposal \\
        9 & V4 & 227 & 0.005 & Cross Contamination Prevention \\
        10 & V1 & 219 & 0.005 & Food Source, Spoilage, Labels \\
        11 & V13 & 90 & 0.002 & No Rodents, Insects, or Animals \\
        12 & V7 & 52 & 0.001 & Wash and Rinse Cycle Temperature \\
        13 & V14 & 37 & 0.001 & Previous Serious Violation Corrected \\
        14 & V5 & 8 & 0.000 & Personnel with Infections Restricted \\
        \hline
    \end{tabularx}
    \caption{Frequency of critical violations in canvass inspections from the public dataset (January 2010 to June 2018).}
    \label{table:training_violations}
\end{table}

\subsubsection{The Target Variable}

Until July 2018, the Chicago food code included three levels of violations: 14 critical violations, i.e. issues that pose immediate health hazards; 15 serious violations, i.e. issues that represent potential health hazards; and 16 minor violations, i.e. issues that do not pose an immediate threat to public health \cite{healthychicago}. Critical violations, while relatively infrequent, represent immediate health hazards. Table \ref{table:training_violations} shows a description of these critical violations and their frequencies and rates in the public dataset.

Table \ref{table:training_violations} shows that even the most common critical violation citation occurs only in less than 10\% of canvass inspections. CDPH model uses a single binary target label indicating whether or not at least one of the 14 critical violation will be cited and does not distinguish between different kinds of critical violations. In the model training data, 14.1\% of the 17,075 instances commit at least one critical violation.

\subsubsection{Predictor Variables}

The inspections are described through 16 features. Ten of the variables contain contextual and environmental data about the inspection, while the remaining six variables are sanitarian cluster variables that indicate which sanitarian group last inspected the restaurant. These features are: 
\begin{packed_enum}
    \item \textit{Past Serious Violation}: Whether or not the food establishment was cited for a serious violation in the last inspection.
    \item \textit{Past Critical Violation}: Whether or not the food establishment was cited for a critical violation in the last inspection.
    \item \textit{Time Since Last Inspection}: The time elapsed since the establishment's last canvass inspection, in fractional years.
    \item \textit{Age at Inspection}: Whether or not the food establishment's business license was more than four years old at time of inspection.
    \item \textit{Alcohol License}: Whether or not the establishment has a license for alcohol consumption on premises.
    \item \textit{Tobacco License}: Whether or not the food establishment has a license to sell tobacco.
    \item \textit{Daily High Temperature}: The daily high temperature, in Fahrenheit, on the day of inspection.
    \item \textit{Intensity of Local Burglaries}: The number of burglaries near the food establishment over the last 90 days, smoothed with kernel density estimates.
    \item \textit{Intensity of Local Sanitation Complaints}: The number of sanitation complaints near the food establishment over the last 90 days, smoothed with kernel density estimates.
    \item \textit{Intensity of Local Garbage Cart Requests}:  The number of requests to replace missing or damaged garbage carts by residents near the food establishment over the last 90 days, smoothed with kernel density estimates.
    \item And 6 additional features relating to the sanitarian who performed the last inspection. We discuss these features in detail next.
\end{packed_enum}

\paragraph{Sanitarian Features}

The model used by CDPH includes a separate variable for each sanitarian. To protect the identity of each sanitarian, CDPH grouped the sanitarians into six clusters based on their coefficients in the full model, gave each cluster a color name, and provided the cluster name of the sanitarian as the predictive feature. Table \ref{table:san_clusters} shows the cluster names, the number of inspections they performed in the training data, and their hit rates (the frequency of inspections that found a critical violation).


\begin{table}[ht]
    \centering
    \begin{tabular}{|l|l|l|}
        \hline
        \textbf{Cluster} & \textbf{Inspections} & \textbf{Hit Rate} \\
        \hline
        Purple & 1174 & 0.406 \\
        Blue & 2897 & 0.265 \\
        Orange & 3769 & 0.136 \\
        Green & 4595 & 0.095 \\
        Yellow & 2762 & 0.058 \\
        Brown & 1878 & 0.024 \\
        \hline
    \end{tabular}
    \caption{The number of inspections and the critical violation hit rates of sanitarian clusters in the training data.}
    \label{table:san_clusters}
\end{table}

\subsection{Predictive Model}

CDPH uses a logistic regression model with 16 predictor variables and the target variable as described above. Table \ref{table:coefs} shows the coefficients of the model for each of the predictive variables. A positive weight indicates an increased likelihood of committing a critical violation, whereas a negative weight indicates the reverse. 

\begin{table}[ht]
    \begin{tabular}{|l|l|}
        \hline
        \textbf{Feature} & \textbf{Coefficient} \\
        \hline
        Purple Sanitarian Cluster & 1.555 \\
        Blue Sanitarian Cluster & 0.950 \\
        Orange Sanitarian Cluster & 0.202 \\
        Green Sanitarian Cluster & -0.244 \\
        Yellow Sanitarian Cluster & -0.697 \\
        Brown Sanitarian Cluster & -1.306 \\
        Past Serious Violation & 0.302 \\
        Past Critical Violation & 0.427 \\
        Time Since Last Inspection & 0.097 \\
        Age at Inspection & -0.164 \\
        Alcohol License & 0.411 \\
        Tobacco License & 0.171 \\
        Daily High Temperature & 0.005 \\
        Intensity of Local Burglaries & 0.002 \\
        Intensity of Local Sanitation Complaints & 0.002 \\
        Intensity of Local Garbage Cart Requests & -0.004 \\
        \hline
    \end{tabular}
    \caption{Logistic regression coefficients for each feature in the CDPH model.}
    \label{table:coefs}
\end{table}

The most influential features according to coefficient magnitudes are the inspector clusters, specifically the purple and brown clusters. The purple cluster, the cluster that had the highest hit rate (Table \ref{table:san_clusters}) has a high positive weight and the brown cluster, the cluster that had the lowest hit rate, has a large negative weight. We analyze this result further in the analysis section (Section \ref{sec.analysis}).

The extant public health literature discusses a number of these predictors. For example, veteran restaurant owners tend to have greater knowledge of their municipal food code, supporting the negative weight for age at inspection \cite{davies14a}. It has also been shown that inspection frequency supports the positive weight for time since last inspection \cite{bader78a}.

The environmental features, such as intensities of the local burglaries, sanitarian complaints, and garbage cart requests, have the lowest weights. These variables were not scaled, which might partially explain why they have smaller coefficients.

\begin{figure}[hb]
	\centering
	\begin{subfigure}[t]{\columnwidth}
		\centering
        \includegraphics[width=\columnwidth]{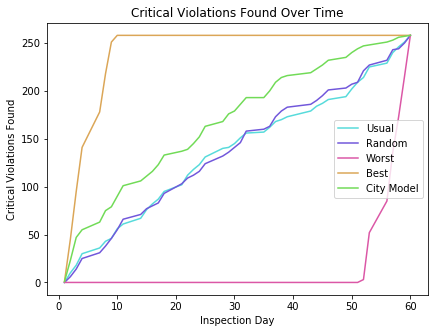}
        \caption{Cumulative number of restaurants found with at least one critical violation over time for various simulated schedules.}
        \label{figure:hit_curves}
	\end{subfigure}
	\quad
	\begin{subfigure}[t]{\columnwidth}
		\centering
        \includegraphics[width=\columnwidth]{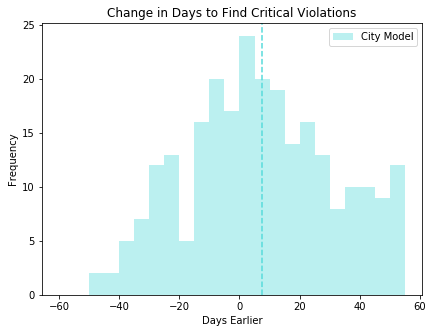}
        \caption{Histogram of reduction in days to inspect restaurants with at least one critical violation using the CDPH model schedule.}
        \label{figure:days_change}
	\end{subfigure}
	\caption{Performance metrics for the food inspection forecasting model.}
\end{figure}

\break
\subsection{Performance Metrics}

CDPH considers three metrics when evaluating simulated canvass inspection schedules:

\begin{enumerate}
    \item Average of reduction in days to inspect restaurants with a critical violation.
    \item Standard deviation of reduction in days to inspect restaurants with a critical violation.
    \item Fraction of restaurants with critical violations visited in the first half of inspections.
\end{enumerate}

The test set includes 1,637 labeled instances. Figure \ref{figure:hit_curves} compares various simulated schedules for inspections in the test set based on the cumulative number of restaurants found with critical violations over time. The series ``Usual`` represents the order of inspections in real life, ``Random`` represents a random ordering of restaurants, ``Worst`` represents a schedule where all restaurants with a critical violation are inspected at the end of the schedule, ``Best`` represents a schedule where all restaurants with a critical violation are inspected at the start of the schedule, and ``City Model`` represents the restaurants scheduled in decreasing order of the model’s predicted probability of committing a critical violation. In the simulated evaluation, CDPH assumes that the same number of canvass inspections can be conducted each day. Figure \ref{figure:days_change} shows the distribution of values for reduction in days to inspect restaurants with at least one critical violation in the simulated schedule using the CDPH model. Note that some restaurants with critical violations are scheduled for a later date than their actual inspection date.

According to these simulated metrics, the CDPH model reduces the time to find restaurants with at least one critical violation by 7.438 days on average with a standard deviation of 25.156 days. The first half of inspections scheduled by the CDPH model includes 69\% of restaurants with at least one critical violation. According to these simulated metrics, the CDPH model accelerates the discovery of critical violations. We further analyze these results in the next section.
\section{Detailed Analysis of the Model and Its Experimental Results}
\label{sec.analysis}

We analyzed the model, its weights, and the results of 50,462 canvass inspections and 17,088 complaint inspections from January 2010 to June 2018 and present five main findings.

\begin{finding}
Using sanitarian clusters as predictor variables unfairly changes predicted risk.
\end{finding}

\begin{figure}[h]
    \centering
    \includegraphics[width=\columnwidth]{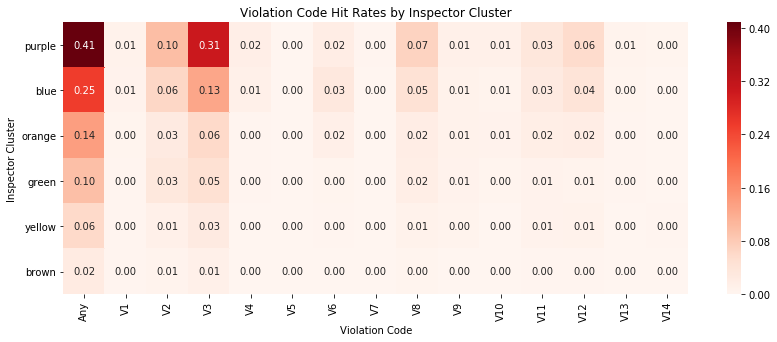}
    \caption{Violation code hit rates by sanitarian cluster for each critical violation code.}
    \label{figure:sanitarian_hit_rates}
\end{figure}

The most influential model feature is the sanitarian cluster who conducted the most recent inspection. Figure \ref{figure:sanitarian_hit_rates} shows the hit rates in Table \ref{table:san_clusters} for each cluster and each critical violation code. The purple cluster appears to have the highest hit rates for V2, V3, V8, and V12. The purple and blue clusters are the only groups of sanitarians where at least one critical violation code has a hit rate higher than 10\%.

The differences in cluster hit rates imply either or both of the following:

\begin{itemize}
    \item Sanitarians vary in ``strictness,'' citing food establishments for critical violations at different frequencies.
    \item Food establishments vary in levels of compliance and underlying risk across sanitarian clusters.
\end{itemize}

In other geographic areas, the extant public health literature shows support for the former claim. Research on food inspections in the Seattle-King County of Washington concludes that sanitarians are consistent in assessing temperature-related violations but not in their overall inspection scores or when assessing combinations of violations \cite{irwin89a}. A study of 167,574 restaurant inspections in Tennessee from January 1993 to April 2000 found high variance among the inspection results of 190 sanitarians who had each conducted 100 inspections or more \cite{jones04a}. In Massachusetts, an independent state audit of local food protection authorities found that inspectors were qualitatively different in their interpretations of standard violations, and that they provided inconsistent follow-up instructions to restaurants \cite{denucci07a}

With regard to the Chicago case, shown in Figure \ref{figure:sanitarian_hit_rates}, ``purple'' sanitarians have the highest average hit rate ($0.406$). The indicator variable for the purple cluster has a coefficient of $1.555$ in the CDPH logistic regression model, meaning that if a food establishment was last inspected by a purple sanitarian, the modeled odds ratio of that food establishment committing a critical violation for the next inspection is $4.735$, even if the food establishment had no critical or serious violations in the previous inspection. In the two-month test set of 1,637 food establishment inspections, the model scheduled all 99 Chicago food establishments last inspected by purple sanitarians ($6.0$\% of the total test set) to be visited among the first 234 inspections ($14.3$\% of the total test set). Of these 99 food establishments, 43 were cited for at least one critical violation ($43.3$\%).

In Chicago, it may be that food establishments inspected previously by purple sanitarians have much higher underlying risk compared to food establishments last inspected by brown sanitarians; however, this variable aggregates compliant food establishments with non-compliant food establishments solely on the basis of the type of sanitarian last inspecting them. Given that food establishments have no control over the sanitarian inspecting them, this information should not factor into their risk score, let alone be the determining factor. 

\break

\begin{finding}
The time invariance assumption may not apply to all food establishments.
\end{finding}

The performance metrics measured by CDPH (please see Section \ref{sec.model} for details) rest on the assumption that critical violations found during the two months of the test set are time invariant, i.e., that any food establishment cited for a violation would also have been cited for a violation had it been inspected on another day in the test period. The time invariance assumption also applies to food establishments not cited for a violation.

\begin{figure}[!hb]
	\centering
	\begin{subfigure}[t]{0.9\columnwidth}
		\centering
		\includegraphics[width=\columnwidth]{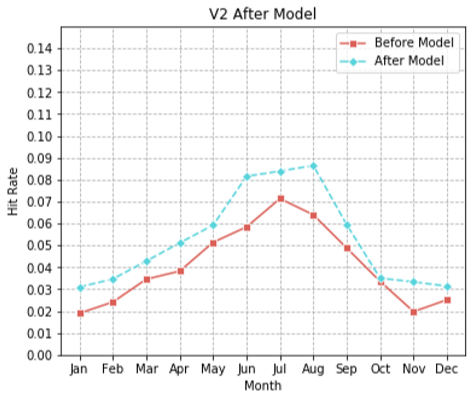}
		\caption{V2 hit rates before and after 2015.}
	\end{subfigure}
	\quad
	\begin{subfigure}[t]{0.9\columnwidth}
		\centering
		\includegraphics[width=\columnwidth]{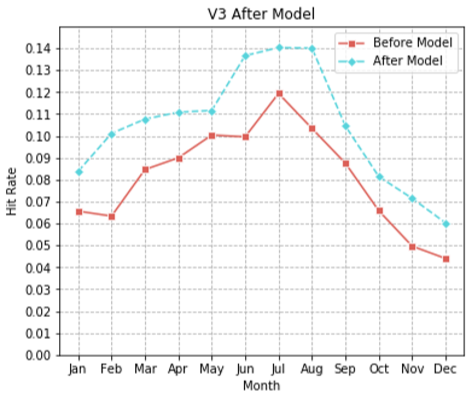}
		\caption{V3 hit rates before and after 2015.}
	\end{subfigure}
	\caption{Mean monthly hit rate for critical violation codes V2 and V3 before and after 2015.}\label{fig.time.invariance}
\end{figure}

If the time invariance assumption does not hold, evaluating the performance of food inspection forecasting models leads to a counterfactual analysis problem: Would a food establishment have been cited for a critical violation if it were inspected on another day? Counterfactual problems are prevalent in machine learning, especially in applications of civic technology. Decisions about bail bonds require judges, or the algorithms that assist them, to predict what a defendant would do if they were released before their trial \cite{kleinberg17a}. Choosing which reports of child abuse to investigate further require hotline staff, or the algorithms that assist them, to predict what would happen to a child’s case if no action were taken \cite{chouldechova18a}

We found that two temperature-related critical violations, V2 and V3, appear to be more frequent in canvass inspections during the summer months, shown in Figures \ref{fig.time.invariance} (a) and (b). On this basis, we analyzed 8,783 canvass inspections of the 51 most common chain restaurants, controlling for each restaurant’s associated chain. We found that monthly average temperature was positively associated with V2 and V3.

\begin{finding}
Increases in critical violation citation rates may \underline{not} be due to the model.
\label{finding.model.success}
\end{finding}

The purpose of using the model is to catch critical violations earlier. However, since we do not know when a food establishment would have been inspected if the model were not in use, we cannot determine how early a violation was identified. Instead, we compare hit rates\footnote{Even though we measure hit rates here, please see Finding \ref{finding.hit.rate} for why hit rate might not be an ideal metric.} before and after January 2015, when CDPH started using the model.

Among canvass inspections, we found that the hit rates of critical violations V2, V3, V6, V11, and V12 increased after 2015, while the hit rate of V4 decreased after 2015.

We assume there is a relationship between canvass and complaint inspections: if the model helps CDPH catch violations earlier, then there will be fewer complaints related to those issues. Thus, if the model does indeed identify critical violations earlier than previously, we would expect the following:

\begin{packed_item}
    \item Monthly hit rates of canvass inspections \textbf{remain the same or increase} after 2015
    \item Monthly hit rates of complaint inspections \textbf{remain the same or decrease} after 2015
\end{packed_item}

\begin{figure*}[ht]
    \centering
    \includegraphics[width=\textwidth]{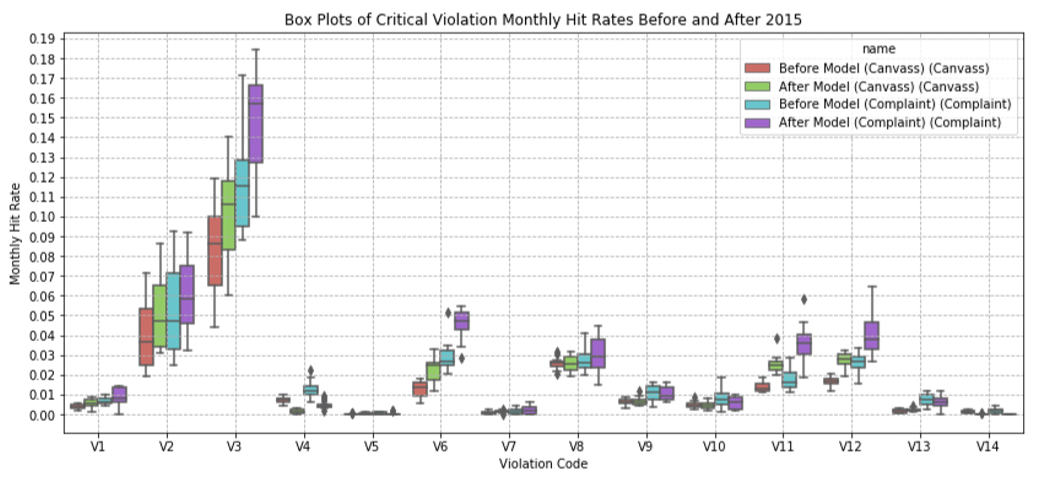}
    \caption{Boxplots of monthly hit rates for each critical violation code, in canvass and complaint inspections, both before and after the model deployment in 2015.}
    \label{fig.monthly.hit.rates}
\end{figure*}

Contrary to our expectations, and shown in Figure \ref{fig.monthly.hit.rates}, the hit rates of both canvass and complaint inspections increased after 2015. We offer two alternative explanations for the changes in critical violation hit rates:
\begin{packed_item}
    \item Food establishment openings and closings have changed the distribution of critical violations.
    \item Inspectors have become stricter on certain critical violation codes due to improved training.
\end{packed_item}

Based on this, it is not clear whether or not the deployment of the model has been effective in discovering violations earlier.

\begin{finding}
Violation hit rate is not an ideal metric for inspection forecasting.
\label{finding.hit.rate}
\end{finding}

Routine food inspections serve to protect employees and patrons from health hazards and to educate food establishments on how to comply with the food code. However, hit rates do not capture improvements to food safety across the city or in individual food establishments. In the extant literature, there is disagreement regarding whether or not compliance with food inspections reduces the risk of food-borne illness outbreaks \cite{irwin89a,cruz01a,jones04a}. Increased hit rate can be interpreted both as a success and as a failure. It is a success because the food protection program identifies violations. However, it is also a failure because inspections are not achieving their primary purpose: to reduce violations and protect the public.

Moreover, researchers warn that machine learning applications in government should not predict an outcome that the agency making the prediction controls, otherwise staff may reinforce a feedback loop \cite{chouldechova18a,ensign18a}. The CDPH model does not predict whether or not a food establishment should be inspected but rather it is simply used to prioritize the inspections; hence, all establishments are inspected, lowering the risk of feedback loops. However, even though sanitarians do not know a food establishment’s predicted risk score, if a food establishment appears earlier in their schedule, they may realize that it has been prioritized and be more strict in their evaluation, fulfilling the predicted outcome.

Finally, hit rates may not offer suitable comparisons of public health interventions. CDPH experimented with FINDER, a system that flags food establishments by tracking symptoms of food-borne illness in search and location data from Google users in Chicago \cite{sadilek18a}. Between November 2016 and March 2017, FINDER achieved a hit rate of 52.3\% while canvass inspections had a hit rate of 22.7\%. FINDER appears superior by this measure, but it only led to inspections of 132 food establishments, while there were 9,495 canvass inspections over the same period. The CDPH model must schedule all food establishments, so its hit rate is naturally lower. More importantly, and not captured by these respective hit rates, FINDER is a reactive technique, while routine inspections are proactive in terms of addressing health-related concerns.

\begin{finding}
The feature set is not sufficiently rich to predict food safety issues.
\end{finding}

Aside from the sanitarian clusters, the model uses ten other variables as predictors of critical violations (please see Table \ref{table:coefs}).
The extant public health literature discusses a number of these predictors. For example, tenured restaurant owners tend to have greater knowledge of the food code, supporting the negative weight for age at inspection \cite{davies14a}, and it has been shown that inspection frequency supports the positive weight for time since last inspection \cite{bader78a}.

Given that hazard analysis and critical control points (HACCP) and microbiological tests provide much more granular information about food protection risk factors \cite{bryan81a,kassa01a}, we recommend the inclusion of features related to equipment, training, storage, preparation, display, and service. Based on the violation narratives from sanitarians, we imagine new features might capture information such as:

\begin{packed_item}
    \item Ingredients of interest used by the food establishment (e.g., cheese, lettuce, chicken, fish)
    \item Whether or not the food establishment stores food in a display before being served directly to the customer
    \item Time since last pest control service visit
\end{packed_item}

We also identified a positive association between certain food chains and particular violation codes. With guidance from subject matter experts and an extensive qualitative coding process, additional features can be identified that could explain additional differences between food chains as well as the reasons that certain chains are more likely to be associated with specific food code violations.

\section{Discussion}
\label{sec.discussion}

We reflect on several themes from the food inspection forecasting project that are relevant to others who use AI in government.

\subsection{Open Source}

The City of Chicago hosts the open source food inspection model on GitHub, a web-based service for sharing data and code, providing documentation, triaging software bugs, tracking releases, and discussing improvements to the project remotely. Aside from the additional benefits of developing open source software (e.g., cost-effectiveness, community support, additional security, etc.), the sharing of data and code on GitHub enabled outside researchers such as us to scrutinize the performance and underlying assumptions of the model. This scrutiny is important because problems faced in the food inspections project may represent greater risks to other government services powered by artificial intelligence.

We applaud CDPH staff for these efforts, and we encourage other public sector and government agencies to do the same. Even when one has the best intentions in developing and deploying AI tools (e.g., being conscious of bias and fairness, feedback loops, counter-factual analysis, etc.), open-sourcing the tool facilitates public scrutiny and the addressing of potential problems with the AI tool. 

\subsection{Pragmatic Privacy}

Informing the public about which restaurants are in compliance with food safety regulations is a core goal of a food inspection program \cite{jones04a}. Provisioning open source models and data aligns with this purpose so that citizens may know which restaurants are safe and so that retail food establishments can understand how they are being evaluated.

However, canvass inspections must be unannounced and CDPH cannot release information about even the date ranges for inspection cycles. At the same time, CDPH anonymizes the activities of its sanitarians, releasing data about them only as clustered groups of individual sanitarians. The opacity of these two components exemplifies the pragmatic privacy of the open source model: it allows third-party users to study a model nearly identical to the one used in production without being able to perfectly predict its schedule or identify precisely which sanitarians will be sent to a particular restaurant at some point in the future.

Based on our analysis, we recommend that future work study the time invariance assumption and differences in critical violation hit rates between sanitarians. Both topics hinge on data that is not publicly available, but this is not necessarily a roadblock. External collaborators can analyze public data to help prioritize areas for improvement or demonstrate approaches on simulated data so that authorized representatives can implement the methodology on the private data.

\subsection{Collaborating with Government}

Digital tools can be convenient, but we found in-person meetings with stakeholders crucial to successful collaboration. City of Chicago employees have a strong presence at ChiHackNight, a civic technology meetup in Chicago that served as the initial venue for us to interact with both government employees and citizen contributors who had previously or currently worked on the food inspection forecasting project. Collaboration in this less formal, public setting facilitated the scheduling of our follow-up meetings with CDPH to share findings and offer recommendations.

In-person meetings make it easier for government stakeholders to provide non-obvious context. For example, as shown in Finding \ref{finding.model.success}, we observed increases in the monthly average hit rates of both canvass and complaint inspections, and we were able to suggest directly to government stakeholders that the increase was caused by other factors, such as the opening and closing of restaurants since 2015 or increased strictness in violation citations due to improved sanitarian training programs. CDPH team members took particular interest in Figure \ref{fig.monthly.hit.rates} because they do not currently compare hit rates in that way. After taking time to ask follow-up questions, they shared that the change in hit rates could be explained by changes the food inspection program made based on feedback from the state department of public health.

\subsection{Changing Policy Context}

In July 2018, the City of Chicago implemented an updated food code with changes to the violation structure and provided public guidance on the open data portal \cite{food_code_change_summary}. This explicit schedule for the sunsetting of the old food code allowed us to conduct a hindsight analysis on a static dataset. However, the policy change also creates dynamic elements that deserve attention, namely whether or not the current food inspection model can be generalized to the new food code, and how governments can evaluate the impact of policy changes on AI early on when data is limited.

AI solutions are typically introduced into an existing policy context and are evaluated for improved performance and maintenance of the policy goals. In the case of Chicago's food inspection forecasting model, however, it is the incumbent policy while new policies are being introduced. CDPH could retrain the model on the results of inspections under the new food code. However, given that only one year of results is currently available and that some restaurants are inspected only once every 1-2 years, there might not be enough data to model past critical/serious violations or to accurately represent the distribution of Chicago restaurants. In this case, CDPH should match former violation codes with new violation codes to allow historical inspection results to be used as training data, with the most recent year being used for validation.

One improvement the new food code offers is that violations will be explicitly marked as in or out of compliance, which should improve consistency across sanitarians and improve the quality of the data for machine learning \cite{food_code_change_summary}.
\section{Conclusion}
\label{sec.conclusion}

With a limited number of sanitarians, the Chicago Department of Public Health (CDPH) uses the food inspection forecasting model as a way to prioritize food establishment inspections and catch critical food code violations earlier. CDPH made the code, the model, and the data publicly available on GitHub, which we analyzed in this paper.

Our investigation resulted in several findings and recommendations. We found that the increased hit rates by the sanitarians might not be due to the model, that using sanitarian information as a predictive variable unfairly changes the predicted risk, that the time invariance assumption might not hold, that the hit rate is not an ideal metric for evaluating model success, and that feature set is not sufficiently rich enough to predict violations at the establishment level.

We recommended CDPH to investigate differences in violation citations across sanitarians and stop using sanitarian cluster as a predictor because it unfairly increases the risk scores of food establishments, even if they passed their last inspection. We further recommended that new metrics of success, instead of hit rates, should be adopted, and creation and use of additional features should be investigated. 

\section*{Acknowledgments}
We thank Raed Mansour, Director, Office of Innovation, Chicago Department of Public Health and Gene Leynes, Data Scientist, Chicago Department of Innovation and Technology for many productive conversations about this project. Thanks to their hard work in maintaining this open source project, researchers and citizens outside government can engage in efforts to improve food safety in Chicago.

Mustafa Bilgic's work is partially supported by the National Science Foundation CAREER award no.~1350337.

\bibliography{pubs}
\bibliographystyle{aaai}
\end{document}